\def\apj{{ApJ}}

\def\IJMODPHYS{{Int. J. Mod. Phys.}}
\def\JETP{{JETP}}
\def\JETPL{{JETP Lett.}}
\def\JHEP{{JHEP}}
\def\MPL{{Mod. Phys. Lett.}}
\def\NC{{Nuovo Cim.}}
\def\NP{{Nucl. Phys.}}
\def\PAstZh{{Pis'ma Astron. Zh.}}
\def\PHYREP{{Phys. Rep.}}
\def\PL{{Phys. Lett.}}
\def\PRD{{Phys. Rev. D}}
\def\PRL{{Phys. Rev. Lett.}}
\def\PS{{Physica Scripta}}
\def\PTRSLA{{Phil. Trans. R. Soc. Lond. A}}
\def\PZhETF{{Pis'ma Zh. Eksp. Teor. Fiz.}}
\def\SovAstroL{{Sov. Astron. Lett.}}
\def\ZhETF{{Zh. Eksp. Teor. Fiz.}}

\def\gta{\mathrel{\raise 0.1em \hbox{$>$} \hskip -0.8em \lower 0.4em
   \hbox{$\sim$}}}
\def\lta{\mathrel{\vcenter{\vbox{\offinterlineskip \hbox{$<$}
     \vskip 0.2 pt \hbox{$\sim$}}}}}
\def\tot{{\rm tot}}

\makeatletter
\let\eqnnumsav=\@eqnnum
\def\forcedeqno#1{\global\def\@eqnnum{(\thechapter.#1)}%
   \xdef\@currentlabel{\thechapter.#1}}
\def\clearforcedeqno{\let\@eqnnum=\eqnnumsav}
\def\decreqcounter#1{\addtocounter{equation}{-#1}\ignorespaces}
\makeatother

\documentclass[cup5b]{caps}
\usepackage{graphicx}
\usepackage{amssymb}
\usepackage{ociwsymp2}   

\HeadText{A. H. Guth}  

\begin{document}

\pagenumbering{arabic}

\author[]{ALAN H. GUTH\\Massachusetts Institute of Technology}

\chapter{Inflation}

\begin{abstract}

The basic workings of inflationary models are summarized, along
with the arguments that strongly suggest that our universe is the
product of inflation.  I describe the quantum origin of density
perturbations, giving a heuristic derivation of the scale
invariance of the spectrum and the leading corrections to scale
invariance.  The mechanisms that lead to eternal inflation in
both new and chaotic models are described.  Although the infinity
of pocket universes produced by eternal inflation are
unobservable, it is argued that eternal inflation has real
consequences in terms of the way that predictions are extracted
from theoretical models.  Although inflation is generically
eternal into the future, it is not eternal into the past: it can
be proven under reasonable assumptions that the inflating region
must be incomplete in past directions, so some physics other than
inflation is needed to describe the past boundary of the
inflating region. The ambiguities in defining probabilities in
eternally inflating spacetimes are reviewed, with emphasis on the
youngness paradox that results from a synchronous gauge
regularization technique.

\end{abstract}

\section{Introduction}

I will begin by summarizing the basics of inflation, including a
discussion of how inflation works, and why many of us believe
that our universe almost certainly evolved through some form of
inflation.  This material is mostly not new, although the
observational evidence in support of inflation has recently
become much stronger.  Since observations of the cosmic microwave
background (CMB) power spectrum have become so important, I will
elaborate a bit on how it is determined by inflationary models. 
Then I will move on to discuss eternal inflation, showing how
once inflation starts, it generically continues forever, creating
an infinite number of ``pocket'' universes.  If inflation is
eternal into the future, it is natural to ask if it can also be
eternal into the past.  I will describe a theorem by Borde,
Vilenkin, and me (Borde, Guth, \& Vilenkin 2003) which shows
under mild assumptions that inflation cannot be eternal into the
past, and thus some new physics will be necessary to explain the
ultimate origin of the universe.

\section{How Does Inflation Work?}

The key property of the laws of physics that makes inflation
possible is the existence of states with negative pressure.  The
effects of negative pressure can be seen clearly in the Friedmann
equations,
\begin{eqnarray}
  \xdef\eq{\arabic{equation}}
  \forcedeqno{\eq a}\label{eq:1a}
  \skew{-3} \ddot a(t) &=& - {4 \pi \over 3} G (\rho + 3 p) a \\
  \forcedeqno{\eq b}\label{eq:1b}
  H^2 &=& {8 \pi \over 3} G \rho - {k \over a^2} \\
  \noalign{\hbox{and}}
  \forcedeqno{\eq c}\label{eq:1c}
  \dot \rho &=& - 3 H (\rho + p) \ ,
\end{eqnarray}
\clearforcedeqno  
\decreqcounter{2} 
where
\begin{equation}
  H = {\dot a \over a} \ . 
  \label{eq:2}
\end{equation}
Here $\rho$ is the energy density, $p$ is the pressure, $G$ is
Newton's constant, an overdot denotes a derivative with respect
to the time $t$, and throughout this paper I will use units for
which $\hbar = c = 1$.  The metric is given by the
Robertson-Walker form,
\begin{equation}
  ds^2 = - d t^2 + a^2(t) \, \left\{ {d r^2 \over 1 - k r^2} + r^2 (
     d \theta^2 + \sin^2 \theta \, d \phi^2 \right\} \ ,
  \label{eq:3}
\end{equation}
where $k$ is a constant which by rescaling $a$ can always be
taken to be 0 or $\pm$ 1.

Eq.~(\ref{eq:1a}) clearly shows that a positive pressure
contributes to the decceleration of the universe, but a negative
pressure can cause acceleration.  Thus, a negative pressure
produces a repulsive form of gravity.

Furthermore, the physics of scalar fields makes it easy to
construct states of negative pressure, since the energy-momentum
tensor of a scalar field $\phi(x)$ is given by
\begin{equation}
  T^{\mu \nu} = \partial^\mu \phi \partial^\nu \phi - g^{\mu \nu}
     \left[ {\textstyle {1 \over 2}} \partial_\lambda \phi
     \partial^\lambda \phi + V(\phi) \right] \ ,
  \label{eq:4}
\end{equation}
where $g^{\mu\nu}$ is the metric, with signature $(-1,1,1,1)$,
and $V(\phi)$ is the potential energy density.  The energy
density and pressure are then given by
\begin{eqnarray}
  \rho &=& T^{00} = {\textstyle {1 \over 2}} \dot \phi^2 +
     {\textstyle {1 \over 2}} (\nabla_i \phi)^2  + V(\phi) \ ,
  \label{eq:5}\\ 
  p &=& {\textstyle {1 \over 3}} \sum_{i=1}^3 T_{ii} =
     {\textstyle {1 \over 2}} \dot \phi^2 - {\textstyle {1
     \over 6}} (\nabla_i \phi)^2 - V(\phi) \ .
  \label{eq:6}
\end{eqnarray}
Thus, any state which is dominated by the potential energy of a
scalar field will have negative pressure.  

Alternatively, one can show that any state which has an energy
density that cannot be easily lowered must have a negative
pressure.  Consider, for example, a state for which the energy
density is approximately equal to a constant value $\rho_f$. 
Then, if a region filled with this state of matter expanded by an
amount $dV$, its energy would have to increase by
\begin{equation}
  d U = \rho_f \, d V \ . 
  \label{eq:7}
\end{equation}
This energy must be supplied by whatever force is causing the
expansion, which means that the force must be pulling against a
negative pressure.  The work done by the force is given by
\begin{equation}
  dW = - p_f \, d V \ ,
  \label{eq:8}
\end{equation}
where $p_f$ is the pressure inside the expanding region. 
Equating the work with the change in energy, one finds
\begin{equation}
  p_f = - \rho_f \ ,
  \label{eq:9}
\end{equation}
which is exactly what Eqs.~(\ref{eq:5}) and (\ref{eq:6}) imply for
states in which the energy density is dominated by the potential
energy of a scalar field.  (One can derive the same result from
Eq.~(\ref{eq:1c}), by considering the case for which $\dot \rho =
0$.)

In most inflationary models the energy density $\rho$ is
approximately constant, leading to exponential expansion of the
scale factor.  By inserting Eq.~(\ref{eq:9}) into (\ref{eq:1a}),
one obtains a second order equation for $a(t)$ for which the
late-time asymptotic behavior is given by
\begin{equation}
  a(t) \propto e^{\chi t} \ , \hbox{ where } \chi = \sqrt{{8 \pi
   \over 3} G \rho_f } \ .
  \label{eq:10}
\end{equation}

In the original version of the inflationary theory (Guth 1981),
the state which drove the inflation involved a scalar field in a
local (but not global) minimum of its potential energy function. 
A similar proposal was advanced slightly earlier by Starobinsky
(1979, 1980) as an (unsuccessful) attempt to solve the initial
singularity problem, using curved space quantum field theory
corrections to the energy-momentum tensor to generate the
negative pressure.  The scalar field state employed in the
original version of inflation is called a {\it false vacuum},
since the state temporarily acts as if it were the state of
lowest possible energy density.  Classically this state would be
completely stable, because there would be no energy available to
allow the scalar field to cross the potential energy barrier that
separates it from states of lower energy.  Quantum mechanically,
however, the state would decay by tunneling (Coleman 1977; Callan
\& Coleman 1977; Coleman \& De~Luccia 1980).  Initially it was
hoped that this tunneling process could successfully end
inflation, but it was soon found that the randomness of the
bubble formation when the false vacuum decayed would produce
disastrously large inhomogeneities.  Early work on this problem
by Guth and Weinberg was summarized in Guth (1981), and described
more fully in Guth \& Weinberg (1983).  Hawking, Moss, \& Stewart
(1982) reached similar conclusions from a different point of
view.

\begin{figure}
   \centering
   \includegraphics[width=275pt]{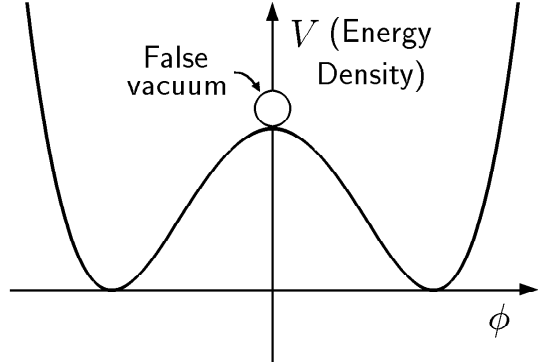}
   \caption{Generic form of the potential for the new inflationary
       scenario.} 
   \label{newinf}
\end{figure}

\begin{figure}
   \centering
   \includegraphics[width=275pt,angle=0]{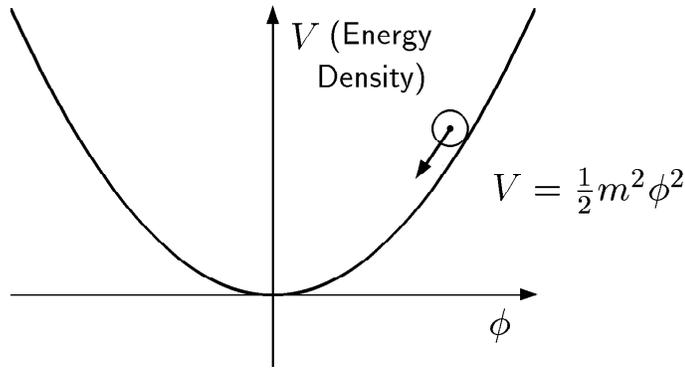}
   \caption{Generic form of the potential for the chaotic inflationary
       scenario.} 
   \label{chaoticinf}
\end{figure}

This ``graceful exit'' problem was solved by the invention of the
new inflationary universe model by Linde (1982a) and by Albrecht
\& Steinhardt (1982).  New inflation achieved all the successes
that had been hoped for in the context of the original version. 
In this theory inflation is driven by a scalar field perched on a
plateau of the potential energy diagram, as shown in
Fig.~\ref{newinf}.  Such a scalar field is generically called the
{\it inflaton}.  If the plateau is flat enough, such a state can
be stable enough for successful inflation.  Soon afterwards Linde
(1983a, 1983b) showed that the inflaton potential need not have
either a local minimum or a gentle plateau: in the scenario he
dubbed {\it chaotic inflation}, the inflaton potential can be as
simple as
\begin{equation} 
   V(\phi)={1 \over 2} m^2 \phi^2,
   \label{eq:11}
\end{equation}
provided that $\phi$ begins at a large enough value so that
inflation can occur as it relaxes.  A graph of this potential
energy function is shown as Fig.~\ref{chaoticinf}.  The evolution
of the scalar field in a Robertson-Walker universe is described
by the general relativistic version of the Klein-Gordon equation,
\begin{equation}
  \ddot \phi + 3 H \dot \phi - {1 \over a^2(t)} \vec \nabla^2
     \phi = - {\partial V \over \partial \phi} \ .
  \label{eq:12}
\end{equation}
For late times the $\vec \nabla^2 \phi$ term becomes negligible,
and the evolution of the scalar field at any point in space is
similar to the motion of a point mass evolving in the potential
$V(x)$ in the presence of a damping force described by the $3 H
\dot \phi$ term.

For simplicity of language, I will stretch the meaning of the
phrase ``false vacuum'' to include all of these cases; that is, I
will use the phrase to denote any state with a large negative
pressure.  

Many versions of inflation have been proposed.  In particular,
versions of inflation that make use of two scalar fields (i.e.,
hybrid inflation (Linde 1991, 1994; Liddle \& Lyth 1993; Copeland
et al. 1994; Stewart 1995) and supernatural inflation (Randall,
Solja\v{c}i\'{c}, \& Guth 1996)) appear to be quite plausible. 
Nonetheless, in this article I will discuss only the basic
scenarios of new and chaotic inflation, which are sufficient to
illustrate the physical effects that I want to discuss.

The basic inflationary scenario begins by assuming that at least
some patch of the early universe was in this peculiar false
vacuum state.  To begin inflation, the patch must be
approximately homogeneous on the scale of $\chi^{-1}$, as defined
by Eq.~(\ref{eq:10}).  In the original papers (Guth 1981; Linde
1982a; Albrecht \& Steinhardt 1982) this initial condition was
motivated by the fact that, in many quantum field theories, the
false vacuum resulted naturally from the supercooling of an
initially hot state in thermal equilibrium.  It was soon found,
however, that quantum fluctuations in the rolling inflaton field
give rise to density perturbations in the universe, and that
these density perturbations would be much larger than observed
unless the inflaton field is very weakly coupled (Starobinsky
1982; Guth \& Pi 1982; Hawking 1982; Bardeen, Steinhardt, \&
Turner 1983).  For such weak coupling there would be no time for
an initially nonthermal state to reach thermal equilibrium. 
Nonetheless, since thermal equilibrium describes a probability
distribution in which all states of a given energy are weighted
equally, the fact that thermal equilibrium leads to a false
vacuum implies that there are many ways of reaching a false
vacuum.  Thus, even in the absence of thermal equilibrium---even
if the universe started in a highly chaotic initial state---it
seems reasonable to assume that some small patches of the early
universe settled into the false vacuum state, as was suggested
for example by Guth (1982). Linde (1983b) pointed out that even
highly improbable initial patches could be important if they
inflated, since the exponential expansion could still cause such
patches to dominate the volume of the universe.  If inflation is
eternal, as I will discuss in Sec.~\ref{eternal}, then the
inflating volume increases without limit, and will presumably
dominate the universe as long as the probability of inflation
starting is not exactly zero.

Once a region of false vacuum materializes, the physics of the
subsequent evolution is rather straightforward. The gravitational
repulsion caused by the negative pressure will drive the region
into a period of exponential expansion.  If the energy density of
the false vacuum is at the grand unified theory scale ($\rho_f
\approx (2 \times 10^{16}\ \hbox{GeV})^4)$, Eq.~(\ref{eq:10})
shows that the time constant $\chi^{-1}$ of the exponential
expansion would be about $10^{-38}$ sec, and that the
corresponding Hubble length would be about $10^{-28}$ cm.
For inflation to achieve its goals, this patch has to expand
exponentially for at least 65 e-foldings, but the amount of
inflation could be much larger than this.  The exponential
expansion dilutes away any particles that are present at the
start of inflation, and also smooths out the metric.  The
expanding region approaches a smooth de Sitter space, independent
of the details of how it began (Jensen \& Stein-Schabes, 1987). 
Eventually, however, the inflaton field at any given location
will roll off the hill, ending inflation.  When it does, the
energy density that has been locked in the inflaton field is
released. Because of the coupling of the inflaton to other
fields, that energy becomes thermalized to produce a hot soup of
particles, which is exactly what had always been taken as the
starting point of the standard big bang theory before inflation
was introduced.  From here on the scenario joins the standard big
bang description.  The role of inflation is to establish
dynamically the initial conditions which otherwise would have to
be postulated. 

The inflationary mechanism produces an entire universe starting
from essentially nothing, so one would naturally want to ask where
the energy for this universe comes from.  The answer is that
it comes from the gravitational field.  The universe did not
begin with this colossal energy stored in the gravitational
field, but rather the gravitational field can supply the energy
because its energy can become negative without bound.  As more
and more positive energy materializes in the form of an
ever-growing region filled with a high-energy scalar field, more
and more negative energy materializes in the form of an expanding
region filled with a gravitational field.  The total energy
remains constant at some very small value, and could in fact be
exactly zero.  There is nothing known that places any limit on
the amount of inflation that can occur while the total energy
remains exactly zero.\footnote{In Newtonian mechanics the energy
density of a gravitational field is unambiguously negative; it
can be derived by the same methods used for the Coulomb field,
but the force law has the opposite sign.  In general relativity
there is no coordinate-invariant way of expressing the energy in
a space that is not asymptotically flat, so many experts prefer
to say that the total energy is undefined.  Either way, there is
agreement that inflation is consistent with the general
relativistic description of energy conservation.}

Note that while inflation was originally developed in the context
of grand unified theories, the only real requirements on the
particle physics are the existence of a false vacuum state, and
the possibility of creating the net baryon number of the universe
after inflation.

\section{Evidence for Inflation}

Inflation is not really a theory, but instead it is a paradigm,
or a class of theories.  As such, it does not make specific
predictions in the same sense that the standard model of particle
physics makes predictions.  Each specific model of inflation
makes definite predictions, but the class of models as a whole
can be tested only by looking for generic features that are
common to most of the models.  Nonetheless, there are a number of
features of the universe that seem to be characteristic
consequences of inflation.  In my opinion, the evidence that our
universe is the result of some form of inflation is very solid. 
Since the term {\it inflation} encompasses a wide range of
detailed theories, it is hard to imagine any reasonable
alternative.\footnote{The cyclic-ekpyrotic model (Steinhardt \&
Turok 2002) is touted by its authors as a rival to inflation, but
in fact it incorporates inflation and uses it to explain why the
universe is so large, homogeneous, isotropic, and flat.}
  
The basic arguments for inflation are as follows:

\leftmargini = 2 \parindent  

\begin{enumerate}
\normalsize  
\item{\it The universe is big}

First of all, we know that the universe is incredibly large: the
visible part of the universe contains about $10^{90}$ particles. 
Since we have all grown up in a large universe, it is easy to
take this fact for granted: of course the universe is big, it's
the whole universe! In ``standard'' Friedmann-Robertson-Walker
cosmology, without inflation, one simply postulates that about
$10^{90}$ or more particles were here from the start.  Many of us
hope, however, that even the creation of the universe can be
described in scientific terms.  Thus, we are led to at least
think about a theory that might explain how the universe got to
be so big.  Whatever that theory is, it has to somehow explain
the number of particles, $10^{90}$ or more.  One simple way
to get such a huge number, with only modest numbers as input,
is for the calculation to involve an exponential.  The
exponential expansion of inflation reduces the problem of
explaining $10^{90}$ particles to the problem of explaining 60 or
70 e-foldings of inflation.  In fact, it is easy to construct
underlying particle theories that will give far more than 70
e-foldings of inflation.  Inflationary cosmology therefore
suggests that, even though the observed universe is incredibly
large, it is only an infinitesimal fraction of the entire
universe.

\item{\it The Hubble expansion}

The Hubble expansion is also easy to take for granted, since we
have all known about it from our earliest readings in cosmology. 
In standard FRW cosmology, the Hubble expansion is part of the
list of postulates that define the initial conditions.  But
inflation actually offers the possibility of explaining how the
Hubble expansion began.  The repulsive gravity associated with
the false vacuum is just what Hubble ordered.  It is exactly the
kind of force needed to propel the universe into a pattern of
motion in which each pair of particles is moving apart with a
velocity proportional to their separation.

\item{\it Homogeneity and isotropy}

The degree of uniformity in the universe is startling.  The
intensity of the cosmic background radiation is the same in all
directions, after it is corrected for the motion of the Earth, to
the incredible precision of one part in 100,000.  To get some
feeling for how high this precision is, we can imagine a marble
that is spherical to one part in 100,000.  The surface of the
marble would have to be shaped to an accuracy of about 1,000
angstroms, a quarter of the wavelength of light.  Although modern
technology makes it possible to grind lenses to
quarter-wavelength accuracy, we would nonetheless be shocked if
we unearthed a stone, produced by natural processes, that was
round to an accuracy of 1,000 angstroms.

The cosmic background radiation was released about 400,000 years
after the big bang, after the universe cooled enough so that the
opaque plasma neutralized into a transparent gas.  The cosmic
background radiation photons have mostly been traveling on
straight lines since then, so they provide an image of what the
universe looked like at 400,000 years after the big bang.  The
observed uniformity of the radiation therefore implies that the
observed universe had become uniform in temperature by that time. 
In standard FRW cosmology, a simple calculation shows that the
uniformity could be established so quickly only if signals could
propagate at about 100 times the speed of light, a proposition
clearly contradicting the known laws of physics. 

In inflationary cosmology, however, the uniformity is easily
explained.  It is created initially on microscopic scales, by
normal thermal-equilibrium processes, and then inflation takes
over and stretches the regions of uniformity to become large
enough to encompass the observed universe and more.

\item{\it The flatness problem}

I find the flatness problem particularly impressive, because of
the extraordinary numbers that it involves.  The problem concerns
the value of the ratio
\begin{equation}
  \Omega_\tot \equiv {\rho_\tot \over \rho_c} \ ,
  \label{eq:13}
\end{equation}
where $\rho_\tot$ is the average total mass density of the
universe and $\rho_c = 3 H^2 / 8 \pi G$ is the critical density,
the density that would make the universe spatially flat.  (In the
definition of ``total mass density,'' I am including the vacuum
energy $\rho_{\rm vac} = \Lambda/ 8 \pi G$ associated with the
cosmological constant $\Lambda$, if it is nonzero.)

For the past several decades there has been general agreement
that $\Omega_\tot$ lies in the range
\begin{equation}
  0.1 \lta \Omega_0 \lta 2 \ ,
  \label{eq:14}
\end{equation}
but for most of this period it was very hard to pinpoint the
value with more precision.  Despite the breadth of this range, the
value of $\Omega$ at early times is highly constrained, since
$\Omega=1$ is an unstable equilibrium point of the standard model
evolution.  Thus, if $\Omega$ was ever {\it exactly} equal to
one, it would remain exactly one forever.  However, if $\Omega$
differed slightly from one in the early universe, that
difference---whether positive or negative---would be amplified
with time.  In particular, it can be shown that $\Omega - 1$
grows as
\begin{equation}
  \Omega - 1 \propto \cases{t &(during the radiation-dominated era)\cr
    t^{2/3} &(during the matter-dominated era)\ .\cr}
  \label{eq:15}
\end{equation}
At $t=1$ sec, for example, when the processes of big bang
nucleosynthesis were just beginning, Dicke and Peebles
(1979) pointed out that $\Omega$ must have equaled one to
an accuracy of one part in $10^{15}$.  Classical cosmology
provides no explanation for this fact---it is simply assumed as
part of the initial conditions.  In the context of modern
particle theory, where we try to push things all the way back to
the Planck time, $10^{-43}$ sec, the problem becomes even more
extreme.  If one specifies the value of $\Omega$ at the Planck
time, it has to equal one to 58 decimal places in order to be
anywhere in the range of Eq.~(\ref{eq:14}) today. 

While this extraordinary flatness of the early universe has no
explanation in classical FRW cosmology, it is a natural
prediction for inflationary cosmology.  During the inflationary
period, instead of $\Omega$ being driven away from one as
described by Eq.~(\ref{eq:15}), $\Omega$ is driven towards one,
with exponential swiftness:
\begin{equation}
  \Omega - 1 \propto e^{-2 H_{\rm inf} t} \ ,
  \label{eq:16}
\end{equation}
where $H_{\rm inf}$ is the Hubble parameter during inflation. 
Thus, as long as there is a sufficient period of inflation,
$\Omega$ can start at almost any value, and it will be driven to
unity by the exponential expansion.  Since this mechanism is
highly effective, almost all inflationary models predict that
$\Omega_0$ should be equal to one (to within about 1 part in
$10^4$).  Until the past few years this prediction was thought to
be at odds with observation, but with the addition of dark energy
the observationally favored value of $\Omega_0$ is now
essentially equal to one.  According to the latest WMAP results
(Bennett et al. 2003), $\Omega_0 = 1.02 \pm 0.02$, in beautiful
agreement with inflationary predictions.

\item{\it Absence of magnetic monopoles}

All grand unified theories predict that there should be, in the
spectrum of possible particles, extremely massive particles
carrying a net magnetic charge.  By combining grand unified
theories with classical cosmology without inflation, Preskill
(1979) found that magnetic monopoles would be produced
so copiously that they would outweigh everything else in the
universe by a factor of about $10^{12}$.  A mass density this
large would cause the inferred age of the universe to drop to
about 30,000 years! Inflation is certainly the simplest known
mechanism to eliminate monopoles from the visible universe, even
though they are still in the spectrum of possible particles.  The
monopoles are eliminated simply by arranging the parameters so
that inflation takes place after (or during) monopole production,
so the monopole density is diluted to a completely negligible
level.

\item{\it Anisotropy of the cosmic microwave background (CMB)
radiation}

The process of inflation smooths the universe essentially
completely, but density fluctuations are generated as inflation
ends by the quantum fluctuations of the inflaton field.  Several
papers emerging from the Nuffield Workshop in Cambridge, UK,
1982, showed that these fluctuations are generically adiabatic,
Gaussian, and nearly scale-invariant (Starobinsky 1982; Guth \&
Pi 1982; Hawking 1982; Bardeen et al. 1983).\footnote{The concept
that quantum fluctuations might provide the seeds for
cosmological density perturbations, which goes back at least to
Sakharov (1965), was pursued in the early 1980s by Lukash (1980a,
1980b), Press (1980, 1981), and Mukhanov \& Chibisov (1981,
1982).  Mukhanov \& Chibisov's papers are of particular interest,
since they considered such quantum fluctuations in the context of
the Starobinsky (1979, 1980) model, now recognized as a version
of inflation.  There is some controversy and ongoing discussion
concerning the historical role of the Mukhanov \& Chibisov
papers, so I include a few comments that the reader can pursue if
interested.  Mukhanov \& Chibisov first discovered that quantum
fluctuations prevent the Starobinsky model from solving the
initial singularity problem.  They then considered the
possibility that the quantum fluctuations are relevant for
density perturbations, and found a nearly scale-invariant
spectrum during the de Sitter phase.  Without any derivation that
I can presently discern, the 1981 paper gives a nearly
scale-invariant formula for the {\it final} density perturbations
after the end of inflation, which is similar but not identical to
the result that was later described in detail in Mukhanov,
Feldman, \& Brandenberger (1992).  In a recent preprint, Mukhanov
(2003) refers to Mukhanov \& Chibisov (1981) as ``the first paper
where the spectrum of inflationary perturbations was
calculated.''  But controversies surrounding this statement
remain unresolved.  Why, for example, were the authors never
explicit about the subtle question of how they calculated the
evolution of the (conformally flat) density perturbations in the
de Sitter phase into the (conformally Newtonian) perturbations
after reheating?  This gap seems particularly evident in the
longer 1982 paper.  And could the Starobinsky model properly be
considered an inflationary model in 1981 or 1982, since at the
time there was no recognition in the literature that the model
could be used to explain the homogeneity, isotropy, or flatness
of the universe?  It was not until later that Whitt (1984) and
Mijic, Morris, \& Suen (1986) established the equivalence between
the Starobinsky model and standard inflation.  After the 1981 and
1982 Mukhanov \& Chibisov papers, the topic of density
perturbations in the Starobinsky model was revisited by a
number of authors, starting with Starobinsky (1983).} 

When my colleagues and I were trying to calculate the spectrum of
density perturbations from inflation in 1982, I never believed
for a moment that it would be measured in my lifetime.  Perhaps
the few lowest moments would be measured, but certainly not
enough to determine a spectrum.  But I was wrong.  The
fluctuations in the CMB have now been measured to exquisite
detail, and even better measurements are in the offing.  So far
everything looks consistent with the predictions of the simplest,
generic inflationary models.  Fig.~\ref{wmappow} shows the
temperature power spectrum and the temperature-polarization
cross-correlation, based on the first year of data of the WMAP
experiment (Bennett et al.\ 2003).  The curve shows the best-fit
``running-index'' $\Lambda$CDM model.  The gray band indicates
one standard deviation of uncertainty due to cosmic variance (the
limitation imposed by being able to sample only one sky). The
underlying primordial spectrum is modeled as a power law
$k^{\, n_s}$, where $n_s=1$ corresponds to scale-invariance.  The
best fit to WMAP alone gives $n_s=0.99 \pm 0.04$.  When WMAP data
is combined with data on smaller scales from other observations
there is some evidence that $n$ grows with scale, but this is not
conclusive.  As mentioned above, the fit gives $\Omega_0 = 1.02
\pm 0.02$.  The addition of isocurvature modes does not improve
the fit, so the expectation of adiabatic perturbations is
confirmed, and various tests for non-Gaussianity have found no
signs of it.
  
\end{enumerate}

\begin{figure}
   \centering
   \includegraphics[width=288pt]{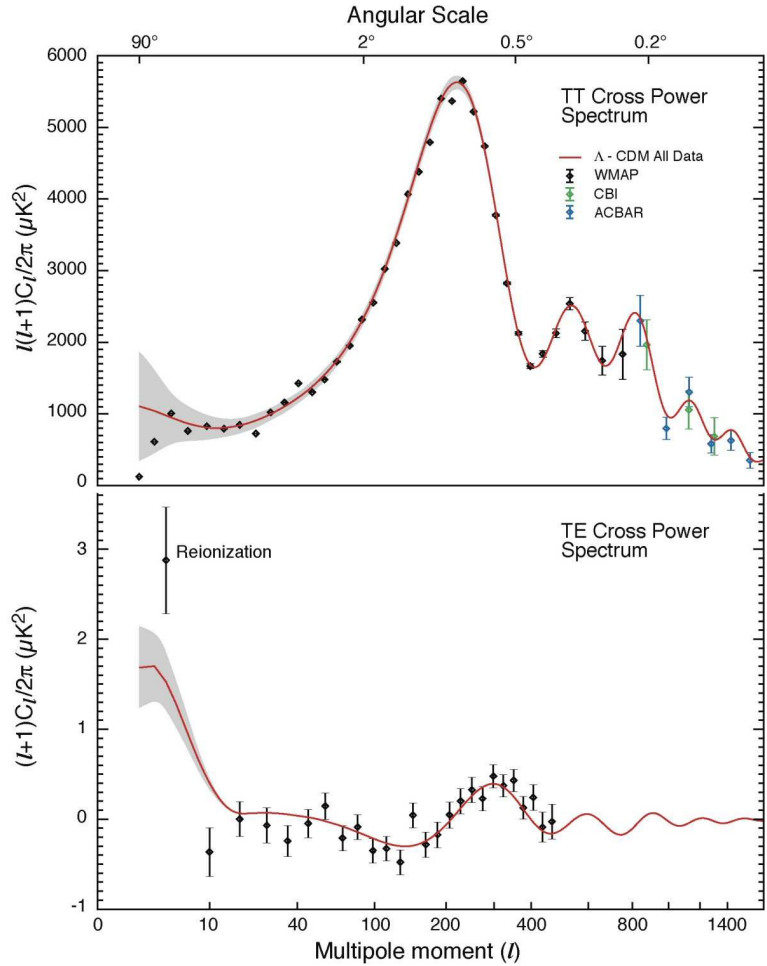}
   \caption{Power spectra of the cosmic background radiation as
     measured by WMAP (Bennett et al. 2003, courtesy of the
     NASA/WMAP Science Team).  The top panel shows the
     temperature anisotropies, and the bottom panel shows the
     correlation between temperature fluctuations and E-mode
     polarization fluctuations.  The solid line is a fit
     consistent with simple inflationary models.}
   \label{wmappow}
\end{figure}

\section{The Inflationary Power Spectrum}
\label{power}

A complete derivation of the density perturbation spectrum
arising from inflation is a very technical subject, so the
interested reader should refer to the Mukhanov et al.\ (1992) or
Liddle \& Lyth (1993) review articles.  However, in this section
I will describe the basics of the subject, for single field slow
roll inflation, in a simple and qualitative way.

For a flat universe ($k=0$) the metric of Eq.~(\ref{eq:3})
reduces to
\begin{equation}
  ds^2 = - dt^2 + a^2(t) d \vec x^2 \ .
  \label{eq:17}
\end{equation}
The perturbations are described in terms of linear perturbation
theory, so it is natural to describe the perturbations in terms
of a Fourier expansion in the comoving coordinates $\vec x$. 
Each mode will evolve independently of all the other modes. 
During the inflationary era the physical wavelength of any given
mode will grow with the scale factor $a(t)$, and hence will grow
exponentially.  The Hubble length $H^{-1}$, however, is
approximately constant during inflation.  The modes of interest
will start at wavelengths far less than $H^{-1}$, and will grow
during inflation to be perhaps 20 orders of magnitude larger than
$H^{-1}$.  For each mode, we will let $t_1$ (``first Hubble
crossing'') denote the time at which the wavelength is equal to
the Hubble length during the inflationary era.  When inflation is
over the wavelength will continue to grow as the scale factor,
but the scale factor will slow down to behave as $a(t) \propto
t^{1/2}$ during the radiation-dominated era, and $a(t) \propto
t^{2/3}$ during the matter-dominated era.  The Hubble length
$H^{-1}= a/\dot a$ will grow linearly with $t$, so eventually the
Hubble length will overtake the wavelength, and the wave will
come back inside the Hubble length.  We will let $t_2$ (``second
Hubble crossing'') denote the time for each mode when the
wavelength is again equal to the Hubble length.  This pattern of
evolution is important to our understanding, because complicated
physics can happen only when the wavelength is smaller than or
comparable to the Hubble length.  When the wavelength is large
compared to the Hubble length, the distance that light can travel
in a Hubble time becomes small compared to the wavelength, and
hence all motion is very slow and the pattern is essentially
frozen in.

Inflation ends when a scalar field rolls down a hill in a
potential energy diagram, such as Figs.~\ref{newinf} or
\ref{chaoticinf}.  Since the scalar field undergoes quantum
fluctuations, however, the field will not roll homogeneously, but
instead will get a little ahead in some places and a little
behind in others.  Hence inflation will not end everywhere
simultaneously, but instead the ending time will be a function of
position: 
\begin{equation}
  t_{\rm end}(\vec x) = t_{\rm end, average} + \delta t (\vec x)
     \ . 
  \label{eq:18}
\end{equation}
Since some regions will undergo more inflation than others, we
have a natural source of inhomogeneities.

Next we need to define a statistical quantity that characterizes
the perturbations.  Letting ${\delta \rho \over \rho} (\vec x,
t)$ describe the fractional perturbation in the total energy
density $\rho$, useful Fourier space quantities can be defined by
\begin{eqnarray}
  \xdef\eq{\arabic{equation}}
  \label{eq:19} 
  \forcedeqno{\eq a}\label{eq:19a}
  \left[ {\delta \tilde \rho \over \rho} (\vec k, t) \right]^2
     &\equiv& {k^3 \over (2 \pi)^3} \int d^3 x \, e^{i \vec k \cdot
     \vec x} \left\langle {\delta \rho \over \rho} (\vec x, t)
     \, {\delta \rho \over \rho} (\vec 0, t) \right\rangle \ , \\
  \forcedeqno{\eq b}\label{eq:19b}
  \left[ {\delta \tilde t} (\vec k) \right]^2
     &\equiv& {k^3 \over (2 \pi)^3} \int d^3 x \, e^{i \vec k \cdot
     \vec x} \left\langle {\delta t} (\vec x)
     \, {\delta t} (\vec 0) \right\rangle \ , 
\end{eqnarray}
\clearforcedeqno  
\decreqcounter{1} 
where the brackets denote an expectation value.

Since the wave pattern is frozen when the wavelength is large
compared to the Hubble length, for any given mode $\vec k$ the
pattern is frozen between $t_1(\vec k)$ and $t_2(\vec k)$.  We
therefore expect a simple relationship between the amplitude of
the perturbation at times $t_1$ and $t_2$, where the perturbation
at time $t_1$ is described by a time offset $\delta t$ in the
evolution of the scalar field, and at $t_2$ it is described by
$\delta \rho / \rho$.  Since we are approximating the problem
with first order perturbation theory, the relationship must be
linear.  By dimensional analysis, the relationship must have the
form 
\begin{equation}
  {\delta \tilde \rho \over \rho} \bigl(\vec k, t_2(k)\bigr) =
     C_1 H(t_1) \, \delta \tilde t(\vec k) \ ,
  \label{eq:20}
\end{equation}
where $C_1$ is a dimensionless constant and $H$ is the only
quantity with units of inverse time that seems to have relevance. 
Of course deriving Eq.~(\ref{eq:20}) and determining the value
of $C_1$ is a lot of work. 

To estimate $\delta \tilde t(\vec k)$, note that we expect its
value to become frozen at about time $t_1(k)$.  If the classical,
homogeneous rolling of the scalar field down the hill is
described by $\phi_0(t)$, then the offset in time $\delta t$ is
equivalent to an offset of the value of the scalar field, 
\begin{equation}
  \delta \phi = - \dot \phi_0 \, \delta t \ .
  \label{eq:21}
\end{equation}
The sign is not very important, but it is negative because
inflation will end earliest ($\delta t < 0$) in regions where the
scalar field has advanced the most ($\delta \phi > 0$, assuming
$\dot \phi > 0$).  $\dot \phi_0$ is in principle calculable by
solving Eq.~(\ref{eq:12}), omitting the spatial Laplacian term. 
Although it is a 2nd order equation, for ``slow roll'' inflation
one assumes that the $\ddot \phi$ term is negligible, so
\begin{equation}
  \dot \phi = - {1 \over 3 H} \, {\partial V \over \partial \phi}
     \ , \qquad \hbox{where} \quad H^2 = {8 \pi \over 3 M_p^2} V
     \ ,
  \label{eq:22}
\end{equation}
where $M_p \equiv 1/\sqrt{G} = 1.22 \times 10^{19}$ GeV is the
Planck mass.  $\delta \phi$ can be estimated by defining the
quantity $\delta \tilde \phi (\vec k, t)$ in analogy to
Eqs.~(\ref{eq:19}), but the quantity on the right-hand side is
just the scalar field propagator of quantum field theory.  One
can approximate $\delta \phi(\vec x, t)$ as a free massless
quantum field evolving in de Sitter space (see for example
Birrell
\& Davies, 1982).  We want to evaluate $\delta \tilde \phi (\vec
k, t)$ for $| \vec k_{\rm physical} | \approx H$.  Again we can
rely on dimensional analysis, since $\phi$ has the units of mass,
and the only relevant quantity with dimensions of mass is $H$. 
Thus $\delta
\tilde \phi \approx H$, and Eqs.~(\ref{eq:20})--(\ref{eq:22})
can be combined to give
\begin{equation}
  {\delta \tilde \rho \over \rho} \bigl(\vec k, t_2(k)\bigr) =
     C_2 \, \left. {H^2 \over \dot \phi_0} \right|_{t_1(\vec k)}
     = C_3 \, \left. {V^{3/2} \over M_p^3 \, V'}
     \right|_{t_1(\vec k)}  \ , 
  \label{eq:23}
\end{equation}
where $C_2$ and $C_3$ are dimensionless constants, and $V' \equiv
\partial V / \partial \phi$.  The entire quantity on the
right-hand side is evaluated at $t_1(\vec k)$, since it is at
this time that the amplitude of the mode is frozen.

Eq.~(\ref{eq:23}) is the key result.  It describes density
perturbations which are nearly scale invariant, meaning that
$\delta \tilde \rho \bigl(\vec k, t_2(k)\bigr)/ \rho $ is
approximately independent of $k$, because typically $V(\phi)$ and
$V'(\phi)$ are nearly constant during the period when
perturbations of observable wavelengths are passing through the
Hubble length during inflation.  Since $\delta \tilde \rho /
\rho$ is measurable and $C_3$ is calculable, one can use
Eq.~(\ref{eq:23}) to determine the value of $V^{3/2}/(M_p^3 V')$. 
Using COBE data, Liddle and Lyth (1993) found
\begin{equation}
  {V^{3/2} \over M_p^3 \, V'} \approx 3.6 \times 10^{-6} \ .
  \label{eq:24}
\end{equation}

While Eq.~(\ref{eq:23}) describes density perturbations that are
nearly scale invariant, it also allows us to express the
departure from scale invariance in terms of derivatives of the
potential $V(\phi)$.  One defines the scalar index $n_s$ by
\begin{equation}
  \left[{\delta \tilde \rho \over \rho} \bigl(\vec k,
     t_2(k)\bigr)\right]^2 \propto k^{\, n_s - 1} \ ,
  \label{eq:25}
\end{equation}
so
\begin{equation}
  n_s - 1 = {d \ln \left[{\delta \tilde \rho \over \rho} \bigl(\vec k,
     t_2(k)\bigr)\right]^2 \over d \ln k} \ .
  \label{eq:26}
\end{equation}
To carry out the differentiation, note that $k$ is related to
$t_1$ by $H = k/\bigl(2 \pi a(t_1)\bigr)$.  Treating $H$ as a
constant since it varies much more slowly than $a$,
differentiation gives $d k / d t_1 = H k$.  Using
Eq.~(\ref{eq:22}) for $d \phi_0 / d t$, one has (Liddle \& Lyth
1992)
\begin{eqnarray}
  n_s &=& 1+ k \, {d t_1 \over d k} \, {d \phi_0 \over d t_1} {d
     \over d \phi} \ln \left[C_3 \, {V^{3/2} \over M_p^3 \,
     V'} \right]^2 \nonumber \\
    &=& 1 + 2 \eta - 6 \epsilon \ ,
  \label{eq:27}
\end{eqnarray}
where
\begin{equation}
  \epsilon = {M_p^2 \over 16 \pi} \, \left( {V' \over V}
     \right)^2 \ , \qquad \eta = {M_p^2 \over 8 \pi} \left( {V''
     \over V} \right) \ .
  \label{eq:28}
\end{equation}
$\epsilon$ and $\eta$ are the now well-known slow-roll parameters
that quantify departures from scale invariance.  (But the reader
should beware that some authors use slightly different
definitions.)  Alternatively, Eqs.~(\ref{eq:22}) can be used to
express $\epsilon$ and $\eta$ in terms of time derivatives of
$H$:
\begin{equation}
  \epsilon = -{\dot H \over H^2} \ , \qquad 
  \eta = -{\dot H \over H^2} - {\ddot H \over 2 H \dot H} \ ,
  \label{eq:29}
\end{equation}
so
\begin{equation}
  n_s = 1 + 4 {\dot H \over H^2} - {\ddot H \over  H \dot H} \ .
  \label{eq:30}
\end{equation}

The above equation can be used to motivate a generic estimate of
how much $n_s$ is likely to deviate from 1.  Since inflation
needs to end at roughly 60 e-folds from the time $t_1(k)$ when
the right-hand side of Eq.~(\ref{eq:23}) is evaluated, we can
take $60 \, H^{-1}$ as the typical time scale for the variation
of physical quantities.  For any quantity $X$ we can generically
estimate that $|\dot X| \sim H X / 60$, so $n_s - 1 \sim \pm {4
\over 60} \pm {1 \over 60}$.  We can conclude that typically
$n_s$ will deviate from 1 by an amount of order 0.1.  Of course
any detailed model will make a precise prediction for the value
of $n_s$.

\section{Eternal Inflation: Mechanisms}
\label{eternal}

The remainder of this article will discuss eternal
inflation---the questions that it can answer, and the questions
that it raises.  In this section I discuss the mechanisms that
make eternal inflation possible, leaving the other issues for the
following sections.  I will discuss eternal inflation first in
the context of new inflation, and then in the context of chaotic
inflation, where it is more subtle. 

In the case of new inflation, the exponential expansion occurs as
the scalar field rolls from the false vacuum state at the peak of
the potential energy diagram (see Fig.~\ref{newinf}) towards the
trough.  The eternal aspect occurs while the scalar field is
hovering around the peak.  The first model of this type was
constructed by Steinhardt (1983), and later that year Vilenkin
(1983) showed that new inflationary models are generically
eternal.  The key point is that, even though classically the
field would roll off the hill, quantum-mechanically there is
always an amplitude, a tail of the wave function, for it to
remain at the top.  If you ask how fast does this tail of the
wave function fall off with time, the answer in almost any model
is that it falls off exponentially with time, just like the decay
of most metastable states (Guth \& Pi 1985).  The time scale for
the decay of the false vacuum is controlled by
\begin{equation}
  m^2 = -\left.{\partial^2 V \over \partial
     \phi^2}\right|_{\phi=0} \ ,
  \label{eq:31}
\end{equation}
the negative mass-squared of the scalar field when it is at the
top of the hill in the potential diagram.  This is an adjustable
parameter as far as our use of the model is concerned, but $m$
has to be small compared to the Hubble constant or else the model
does not lead to enough inflation.  So, for parameters that are
chosen to make the inflationary model work, the exponential decay
of the false vacuum is slower than the exponential expansion. 
Even though the false vacuum is decaying, the expansion outruns
the decay and the total volume of false vacuum actually increases
with time rather than decreases.  Thus inflation does not end
everywhere at once, but instead inflation ends in localized
patches, in a succession that continues ad infinitum.  Each patch
is essentially a whole universe --- at least its residents will
consider it a whole universe --- and so inflation can be said to
produce not just one universe, but an infinite number of
universes.  These universes are sometimes called bubble
universes, but I prefer to use the phrase ``pocket universe,'' to
avoid the implication that they are approximately round.  (While
bubbles formed in first-order phase transitions are round
(Coleman \& De~Luccia 1980), the local universes formed in eternal
new inflation are generally very irregular, as can be seen for
example in the two-dimensional simulation in Fig.~2 of Vanchurin,
Vilenkin, \& Winitzki (2000).)

In the context of chaotic inflationary models the situation is
slightly more subtle.  Andrei Linde (1986a, 1986b, 1990) showed
that these models are eternal in 1986.  In this case inflation
occurs as the scalar field rolls down a hill of the potential
energy diagram, as in Fig.~\ref{chaoticinf}, starting high on the
hill.  As the field rolls down the hill, quantum fluctuations
will be superimposed on top of the classical motion.  The best
way to think about this is to ask what happens during one time
interval of duration $\Delta t = H^{-1}$ (one Hubble time), in a
region of one Hubble volume $H^{-3}$.  Suppose that $\phi_0$ is
the average value of $\phi$ in this region, at the start of the
time interval.  By the definition of a Hubble time, we know how
much expansion is going to occur during the time interval:
exactly a factor of $e$.  (This is the only exact number in
today's talk, so I wanted to emphasize the point.)  That means
the volume will expand by a factor of $e^3$.  One of the deep
truths that one learns by working on inflation is that $e^3$ is
about equal to 20, so the volume will expand by a factor of 20. 
Since correlations typically extend over about a Hubble length,
by the end of one Hubble time, the initial Hubble-sized region
grows and breaks up into 20 independent Hubble-sized regions. 

As the scalar field is classically rolling down the hill, the
classical change in the field $\Delta \phi_{\rm cl}$ during the
time interval $\Delta t$ is going to be modified by quantum
fluctuations $\Delta \phi_{\rm qu}$, which can drive the field
upward or downward relative to the classical trajectory.  For any
one of the 20 regions at the end of the time interval, we can
describe the change in $\phi$ during the interval by
\begin{equation}
  \Delta \phi = \Delta \phi_{\rm cl} + \Delta \phi_{\rm qu} \ . 
  \label{eq:32}
\end{equation}
In lowest order perturbation theory the fluctuation is treated as
a free quantum field, which implies that $\Delta \phi_{\rm qu}$,
the quantum fluctuation averaged over one of the 20 Hubble
volumes at the end, will have a Gaussian probability
distribution, with a width of order $H/2 \pi$ (Vilenkin \& Ford
1982; Linde 1982b; Starobinsky 1982, 1986). There is then always
some probability that the sum of the two terms on the right-hand
side will be positive --- that the scalar field will fluctuate up
and not down.  As long as that probability is bigger than 1 in
20, then the number of inflating regions with $\phi \ge \phi_0$
will be larger at the end of the time interval $\Delta t$ than it
was at the beginning.  This process will then go on forever, so
inflation will never end. 

Thus, the criterion for eternal inflation is that the probability
for the scalar field to go up must be bigger than $1/e^3 \approx
1/20$.  For a Gaussian probability distribution, this condition
will be met provided that the standard deviation for $\Delta
\phi_{\rm qu} $ is bigger than $0.61 |\Delta \phi_{\rm cl}|$.
Using $\Delta \phi_{\rm cl} \approx \dot \phi_{\rm cl} H^{-1}$,
the criterion becomes
\begin{equation}
  \Delta \phi_{\rm qu} \approx {H \over 2 \pi} > 0.61 \, | \dot
     \phi_{\rm cl}| \, H^{-1} \Longleftrightarrow {H^2 \over |\dot
     \phi_{\rm cl}|} > 3.8 \ . 
 \label{eq:33}
\end{equation}
Comparing with Eq.~(\ref{eq:23}), we see that the condition
for eternal inflation is equivalent to the condition that $\delta
\rho/\rho$ on ultra-long length scales is bigger than a number of
order unity.

The probability that $\Delta \phi$ is positive tends to increase
as one considers larger and larger values of $\phi$, so sooner or
later one reaches the point at which inflation becomes eternal. 
If one takes, for example, a scalar field with a potential
\begin{equation}
  V(\phi) = {1 \over 4} \lambda \phi^4 \ ,
  \label{eq:34}
\end{equation}
then the de Sitter space equation of motion in flat
Robertson-Walker coordinates (Eq.~(\ref{eq:17})) takes the form
\begin{equation}
  \ddot \phi + 3 H \dot \phi = - \lambda \phi^3 \ ,
  \label{eq:35}
\end{equation}
where spatial derivatives have been neglected.  In the
``slow-roll'' approximation one also neglects the $\ddot \phi$
term, so $\dot \phi \approx - \lambda \phi^3 / (3 H)$, where the
Hubble constant $H$ is related to the energy density by
\begin{equation}
  H^2 = {8 \pi \over 3} G \rho = {2 \pi \over 3} {\lambda \phi^4
     \over M_p^2} \ .
  \label{eq:36}
\end{equation}
Putting these relations together, one finds that the criterion
for eternal inflation, Eq.~(\ref{eq:33}), becomes
\begin{equation}
  \phi > 0.75 \, \lambda^{-1/6} \, M_p \ . 
 \label{eq:37}
\end{equation}

Since $\lambda$ must be taken very small, on the order of
$10^{-12}$, for the density perturbations to have the right
magnitude, this value for the field is generally well above the
Planck scale.  The corresponding energy density, however, is
given by
\begin{equation}
  V(\phi) = {1 \over 4} \lambda \phi^4 = .079 \lambda^{1/3} M_p^4
     \ ,
  \label{eq:38}
\end{equation}
which is actually far below the Planck scale.

So for these reasons we think inflation is almost always eternal. 
I think the inevitability of eternal inflation in the context of
new inflation is really unassailable --- I do not see how it
could possibly be avoided, assuming that the rolling of the
scalar field off the top of the hill is slow enough to allow
inflation to be successful.  The argument in the case of chaotic
inflation is less rigorous, but I still feel confident that it is
essentially correct.  For eternal inflation to set in, all one
needs is that the probability for the field to increase in a
given Hubble-sized volume during a Hubble time interval is larger
than 1/20.

Thus, once inflation happens, it produces not just one universe,
but an infinite number of universes. 

\section{Eternal Inflation: Implications}
\label{implications}

In spite of the fact that the other universes created by eternal
inflation are too remote to imagine observing directly, I
nonetheless claim that eternal inflation has real consequences in
terms of the way we extract predictions from theoretical models. 
Specifically, there are three consequences of eternal inflation
that I will discuss.

First, eternal inflation implies that all hypotheses about the
ultimate initial conditions for the universe---such as the Hartle
\& Hawking (1983) no boundary proposal, the tunneling proposals
by Vilenkin (1984, 1986, 1999) or Linde (1984, 1998), or the more
recent Hawking \& Turok (1998) instanton --- become totally
divorced from observation. That is, one would expect that if
inflation is to continue arbitrarily far into the future with the
production of an infinite number of pocket universes, then the
statistical properties of the inflating region should approach a
steady state which is independent of the initial conditions. 
Unfortunately, attempts to quantitatively study this steady state
are severely limited by several factors.  First, there are
ambiguities in defining probabilities, which will be discussed
later.  In addition, the steady state properties seem to depend
strongly on super-Planckian physics which we do not understand. 
That is, the same quantum fluctuations that make eternal chaotic
inflation possible tend to drive the scalar field further and
further up the potential energy curve, so attempts to quantify
the steady state probability distribution (Linde, Linde, \&
Mezhlumian, 1994; Garcia-Bellido \& Linde, 1995) require the
imposition of some kind of a boundary condition at large $\phi$. 
Although these problems remain unsolved, I still believe that it
is reasonable to assume that in the course of its unending
evolution, an eternally inflating universe would lose all memory
of the state in which it started.

Even if the universe forgets the details of its genesis, however,
I would not assume that the question of how the universe began
would lose its interest.  While eternally inflating universes
continue forever once they start, they are apparently not eternal
into the past.  (The word {\it eternal} is therefore not
technically correct---it would be more precise to call this
scenario {\it semi-eternal} or {\it future-eternal}.)  The
possibility of a quantum origin of the universe is very
attractive, and will no doubt be a subject of interest for some
time.  Eternal inflation, however, seems to imply that the entire
study will have to be conducted with literally no input from
observation. 

A second consequence of eternal inflation is that the probability
of the onset of inflation becomes totally irrelevant, provided
that the probability is not identically zero. Various authors in
the past have argued that one type of inflation is more plausible
than another, because the initial conditions that it requires
appear more likely to have occurred.  In the context of eternal
inflation, however, such arguments have no significance.  Any
nonzero probability of onset will produce an infinite spacetime
volume.  If one wants to compare two type of inflation, the
expectation is that the one with the faster exponential time
constant will always win.

A corollary to this argument is that new inflation is not dead. 
While the initial conditions necessary for new inflation cannot
be justified on the basis of thermal equilibrium, as proposed in
the original papers (Linde 1982; Albrecht \& Steinhardt 1982), in
the context of eternal inflation it is sufficient to conclude
that the probability for the required initial conditions is
nonzero.  Since the resulting scenario does not depend on the
words that are used to justify the initial state, the standard
treatment of new inflation remains valid.

A third consequence of eternal inflation is the possibility that
it offers to rescue the predictive power of theoretical physics. 
Here I have in mind the status of string theory, or the theory
known as M theory, into which string theory has evolved.  The
theory itself has an elegant uniqueness, but nonetheless it
appears that the vacuum is far from unique (Bousso \& Polchinski
2000; Susskind 2003).  Since predictions will ultimately depend
on the properties of the vacuum, the predictive power of string/M
theory may be limited.  Eternal inflation, however, provides a
possible mechanism to remedy this problem.  Even if many types of
vacua are equally stable, it may turn out that there is one
unique metastable state that leads to a maximal rate of
inflation.  If so, then this metastable state will dominate the
eternally inflating region, even if its expansion rate is only
infinitesimally larger than the other possibilities.  One would
still need to follow the decay of this metastable state as
inflation ends.  It may very well branch into a number of
final low-energy vacua, but the number that are significantly
populated could hopefully be much smaller than the total number
of vacua.  All of this is pure speculation at this
point, because no one knows how to calculate these things. 
Nonetheless, it possible that eternal inflation might help to
constrain the vacuum state of the real universe, perhaps
significantly enhancing the predictive power of M theory.

\section{Does Inflation Need a Beginning?}

If the universe can be eternal into the future, is it possible
that it is also eternal into the past?  Here I will describe a
recent theorem (Borde, Guth, \& Vilenkin 2003) which shows, under
plausible assumptions, that the answer to this question is
no.\footnote{There were also earlier theorems about this issue by
Borde \& Vilenkin (1994, 1996) and Borde (1994), but these
theorems relied on the weak energy condition, which for a perfect
fluid is equivalent to the condition $\rho + p \ge 0$.  This
condition holds classically for forms of matter that are known or
commonly discussed as theoretical proposals.  It can, however, be
violated by quantum fluctuations (Borde \& Vilenkin 1997), and so
the reliability of these theorems is questionable.}

The theorem is based on the well-known fact that the momentum of
an object traveling on a geodesic through an expanding universe
is redshifted, just as the momentum of a photon is redshifted. 
Suppose, therefore, we consider a timelike or null geodesic
extended backwards, into the past.  In an expanding universe such
a geodesic will be blueshifted.  The theorem shows that under
some circumstances the blueshift reaches infinite rapidity (i.e.,
the speed of light) in a finite amount of proper time (or affine
parameter) along the trajectory, showing that such a trajectory
is (geodesically) incomplete.  

\begin{figure}
   \centering
   \includegraphics[width=240pt]{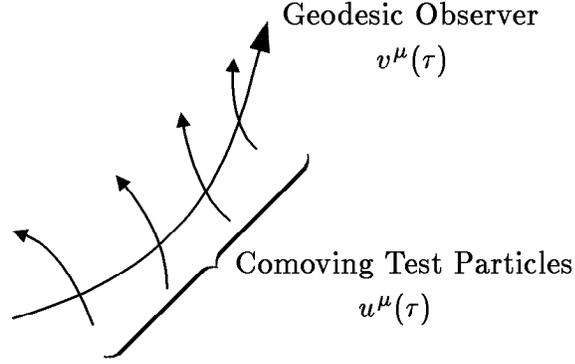}
   \caption{An observer measures the velocity of passing test
       particles to infer the Hubble parameter.}
   \label{geodesic}
\end{figure}

To describe the theorem in detail, we need to quantify what we
mean by an expanding universe.  We imagine an observer whom we
follow backwards in time along a timelike or null geodesic.  The
goal is to define a local Hubble parameter along this geodesic,
which must be well-defined even if the spacetime is neither
homogeneous nor isotropic.  Call the velocity of the geodesic
observer $v^\mu(\tau)$, where $\tau$ is the proper time in the
case of a timelike observer, or an affine parameter in the case
of a null observer.  (Although we are imagining that we are
following the trajectory backwards in time, $\tau$ is defined to
increase in the future timelike direction, as usual.)  To define
$H$, we must imagine that the vicinity of the observer is filled
with ``comoving test particles,'' so that there is a test
particle velocity $u^\mu(\tau)$ assigned to each point $\tau$
along the geodesic trajectory, as shown in Fig.~\ref{geodesic}. 
These particles need not be real --- all that will be necessary
is that the worldlines can be defined, and that each worldline
should have zero proper acceleration at the instant it intercepts
the geodesic observer. 

To define the Hubble parameter that the observer measures at time
$\tau$, the observer focuses on two particles, one that he passes
at time $\tau$, and one at $\tau + \Delta \tau$, where in the end
he takes the limit $\Delta \tau \rightarrow 0$.  The Hubble
parameter is defined by
\begin{equation}
  H \equiv {\Delta v_{\rm radial} \over \Delta r} \ ,
  \label{eq:39}
\end{equation}
where $\Delta v_{\rm radial}$ is the radial component of the
relative velocity between the two particles, and $\Delta r$ is
their distance, where both quantities are computed in the rest
frame of one of the test particles, not in the rest frame of the
observer.  Note that this definition reduces to the usual one if
it is applied to a homogeneous isotropic universe.

The relative velocity between the observer and the test particles
can be measured by the invariant dot product,
\begin{equation}
  \gamma \equiv u_\mu v^\mu \ ,
  \label{eq:40}
\end{equation}
which for the case of a timelike observer is equal to the usual
special relativity Lorentz factor
\begin{equation}
  \gamma = {1 \over \sqrt{1-v^2_{\rm rel}}} \ .
  \label{eq:41}
\end{equation}

If $H$ is positive we would expect $\gamma$ to decrease with
$\tau$, since we expect the observer's momentum relative to the
test particles to redshift.  It turns out, however, that the
relationship between $H$ and changes in $\gamma$ can be made
precise.  If one defines
\begin{equation}
  F(\gamma) \equiv \cases{1/\gamma & for null observers \cr
      {\rm arctanh}(1/\gamma) & for timelike observers \ ,\cr}
  \label{eq:42}
\end{equation}
then
\begin{equation}
  H = {d F(\gamma) \over d \tau} \ .
  \label{eq:43}
\end{equation}

I like to call $F(\gamma)$ the ``slowness'' of the geodesic
observer, because it increases as the observer slows down,
relative to the test particles.  The slowness decreases as we
follow the geodesic backwards in time, but it is positive
definite, and therefore cannot decrease below zero. 
$F(\gamma)=0$ corresponds to $\gamma = \infty$, or a relative
velocity equal to that of light.  This bound allows us to place a
rigorous limit on the integral of Eq.~(\ref{eq:43}).  For
timelike geodesics,
\begin{equation}
  \int^{\tau_f} \, H \, d \tau \le {\rm arctanh}\left({1 \over
     \gamma_f} \right) = {\rm arctanh}\left( \sqrt{1 - v_{\rm
     rel}^2} \right) \ ,
  \label{eq:44}
\end{equation}
where $\gamma_f$ is the value of $\gamma$ at the final time $\tau
= \tau_f$.  For null observers, if we normalize the affine
parameter $\tau$ by $d \tau / d t = 1$ at the final time
$\tau_f$, then
\begin{equation}
  \int^{\tau_f} \, H \, d \tau \le 1 \ .
  \label{eq:45}
\end{equation}
Thus, if we assume an {\it averaged expansion condition}, i.e.,
that the average value of the Hubble parameter $H_{\rm av}$ along
the geodesic is positive, then the proper length (or affine
length for null trajectories) of the backwards-going geodesic is
bounded.  Thus the region for which $H_{\rm av} > 0$ is
past-incomplete. 

It is difficult to apply this theorem to general inflationary
models, since there is no accepted definition of what exactly
defines this class.  However, in standard eternally inflating
models, the future of any point in the inflating region can be
described by a stochastic model (Goncharov, Linde, \& Mukhanov
1987) for inflaton evolution, valid until the end of inflation. 
Except for extremely rare large quantum fluctuations, $H \gta
\sqrt{(8 \pi/3) G \rho_f}$, where $\rho_f$ is the energy density
of the false vacuum driving the inflation.  The past for an
arbitrary model is less certain, but we consider eternal models
for which the past is like the future.  In that case $H$ would be
positive almost everywhere in the past inflating region.  If,
however, $H_{\rm av} > 0$ when averaged over a past-directed
geodesic, our theorem implies that the geodesic is incomplete.

There is of course no conclusion that an eternally inflating
model must have a unique beginning, and no conclusion that there
is an upper bound on the length of all backwards-going geodesics
from a given point.  There may be models with regions of
contraction embedded within the expanding region that could evade
our theorem.  Aguirre \& Gratton (2002, 2003) have proposed a
model that evades our theorem, in which the arrow of time
reverses at the $t=-\infty$ hypersurface, so the universe
``expands'' in both halves of the full de Sitter space.

The theorem does show, however, that an eternally inflating model
of the type usually assumed, which would lead to $H_{\rm av} > 0$
for past-directed geodesics, cannot be complete.  Some new
physics (i.e., not inflation) would be needed to describe the
past boundary of the inflating region.  One possibility would be
some kind of quantum creation event.

One particular application of the theory is the cyclic ekpyrotic
model of Steinhardt \& Turok (2002).  This model has $H_{\rm av}
> 0$ for null geodesics for a single cycle, and since every cycle
is identical, $H_{\rm av} > 0$ when averaged over all cycles. 
The cyclic model is therefore past-incomplete, and requires a
boundary condition in the past.

\section{Calculation of Probabilities in Eternally Inflating
Universes}

In an eternally inflating universe, anything that can happen will
happen; in fact, it will happen an infinite number of times. 
Thus, the question of what is possible becomes trivial---anything
is possible, unless it violates some absolute conservation law. 
To extract predictions from the theory, we must therefore learn
to distinguish the probable from the improbable.

However, as soon as one attempts to define probabilities in an
eternally inflating spacetime, one discovers ambiguities.  The
problem is that the sample space is infinite, in that an
eternally inflating universe produces an infinite number of
pocket universes.  The fraction of universes with any particular
property is therefore equal to infinity divided by infinity---a
meaningless ratio.  To obtain a well-defined answer, one needs to
invoke some method of regularization.  In eternally inflating
universes, however, the answers that one gets depend on how one
chooses the method of regularization.

To understand the nature of the problem, it is useful to think
about the integers as a model system with an infinite number of
entities.  We can ask, for example, what fraction of the integers
are odd.  With the usual ordering of the integers, $1$, $2$, $3$,
\dots, it seems obvious that the answer is $1/2$.  However, the
same set of integers can be ordered by writing two odd integers
followed by one even integer, as in $1,3,\ 2,\ 5,7,\ 4,\ 9,11,\
6\ , \ldots$ .  Taken in this order, it looks like $2/3$ of the
integers are odd.

One simple method of regularization is a cut-off at equal-time
surfaces in a synchronous gauge coordinate system.  Specifically,
suppose that one constructs a Robertson-Walker coordinate system
while the model universe is still in the false vacuum (de Sitter)
phase, before any pocket universes have formed. One can then
propagate this coordinate system forward with a synchronous gauge
condition,\footnote{By a synchronous gauge condition, I mean that
each equal-time hypersurface is obtained by propagating every
point on the previous hypersurface by a fixed infinitesimal time
interval $\Delta t$ in the direction normal to the hypersurface.}
and one can define probabilities by truncating the spacetime
volume at a fixed value $t_f$ of the synchronous time coordinate
$t$.  I will refer to probabilities defined in this way as
synchronous gauge probabilities.

An important peculiarity of synchronous gauge probabilities is
that they lead to what I call the ``youngness paradox''.  The
problem is that the volume of false vacuum is growing
exponentially with time with an extraordinarily small time
constant, in the vicinity of $10^{-37}$ sec. Since the rate at
which pocket universes form is proportional to the volume of
false vacuum, this rate is increasing exponentially with the same
time constant.  This means that for every universe in the sample
of age $t$, there are approximately $\exp\left\{10^{37}\right\}$
universes with age $t - $(1 sec).  The population of pocket
universes is therefore an incredibly youth-dominated society, in
which the mature universes are vastly outnumbered by universes
that have just barely begun to evolve.

Probability calculations in this youth-dominated ensemble lead to
peculiar results, as was first discussed by Linde, Linde, \&
Mezhlumian (1995).  Since mature universes are incredibly rare, it
becomes likely that our universe is actually much younger than we
think, with our part of the universe having reached its apparent
maturity through an unlikely set of quantum jumps.  
These authors considered the expected behavior of the mass
density in our vicinity, concluding that we should find ourselves
very near the center of a spherical low-density region.

Since the probability measure depends on the method used to
truncate the infinite spacetime of eternal inflation, we are not
forced to accept the consequences of the synchronous gauge
probabilities.  A method of calculating probabilities that gives
acceptable answers has been formulated by Vilenkin (1998) and his
collaborators (Vanchurin, Vilenkin, \& Winitzki, 2000; Garriga \&
Vilenkin, 2001).  However, we still do not have a compelling
argument from first principles that determines how probabilities
should be calculated. 

\section{Conclusion}

In this paper I have summarized the workings of inflation, and
the arguments that strongly suggest that our universe is the
product of inflation.  I argued that inflation can explain the
size, the Hubble expansion, the homogeneity, the isotropy, and
the flatness of our universe, as well as the absence of magnetic
monopoles, and even the characteristics of the nonuniformities. 
The detailed observations of the cosmic background radiation
anisotropies continue to fall in line with inflationary
expectations, and the evidence for an accelerating universe fits
beautifully with the inflationary preference for a flat universe. 
Our current picture of the universe seems strange, with 95\% of
the energy in forms of matter that we do not understand, but
nonetheless the picture fits together very well.

Next I turned to the question of eternal inflation, claiming that
essentially all inflationary models are eternal. In my opinion
this makes inflation very robust: if it starts anywhere, at any
time in all of eternity, it produces an infinite number of pocket
universes.  Eternal inflation has the very attractive feature,
from my point of view, that it offers the possibility of allowing
unique (or possibly only constrained) predictions even if the
underlying string theory does not have a unique vacuum.  I
discussed the past of eternally inflating models, concluding that
under mild assumptions the inflating region must have a past
boundary, and that new physics (other than inflation) is needed
to describe what happens at this boundary.  I have also
described, however, that our picture of eternal inflation is not
complete.  In particular, we still do not understand how to
define probabilities in an eternally inflating spacetime. 

The bottom line, however, is that observations in the past few
years have vastly improved our knowledge of the early universe,
and that these new observations have been generally consistent
with the simplest inflationary models.

\section*{Acknowledgments}

This work is supported in part by funds provided by the U.S.
Department of Energy (D.O.E.) under cooperative research
agreement \#DF-FC02-94ER40818.

\end{document}